\documentstyle [12pt,bezier,epsfig]{article}
\begin{document}

\vspace{0.5 cm}

\begin{titlepage}

\begin{flushright}
IFAE-UAB/94-01
\end{flushright}
\begin{flushright}
April 1994
\end{flushright}

\vspace{0.8 cm}

\begin{center}
{\bf\LARGE
About the relevance of the \\
Imaginary components of the effective couplings \\
in the Asymmetry measurements.\\
 }
\end{center}

\vspace{0.8 cm}
\begin{center}
     M.~Mart\'\i nez \\
            and \\
     F.~Teubert\footnote{Research supported by a F.P.I. grant
from the Universitat Aut\`onoma de Barcelona.}  \\
\vspace{0.5 cm}
   Institut de F\'\i sica d'Altes Energies (IFAE)\\
   Universitat Aut\`onoma de Barcelona\\
   {\it Edifici C E-08193 Bellaterra (Barcelona),Spain}
\end{center}

\vspace{0.5 cm}

\begin{center}
{\bf   Abstract}
\vskip 1.0 cm
 The effect coming from imaginary parts of effective couplings
in the $e^+ e^-$ asymmetries is investigated. It is shown that
for the present
level of experimental accuracy, in some asymmetries the imaginary parts
of the effective couplings cannot be neglected and moreover
that the use of different prescriptions on how to handle
them in quoting just real effective couplings from the data
may produce sizable differences. A definition of the real effective
couplings
specifying how to handle the imaginary parts is advocated.
\end{center}

\renewcommand{\baselinestretch}{1.0}

\vspace{0.5cm}
\begin{center}
{\em (Submitted to Z.Phys. C)}
\end{center}
\end{titlepage}
%-----------------------------------------------------------------------
\newpage
%-----------------------------------------------------------------------
\section{Introduction}
%-----------------------------------------------------------------------

    The present accuracy in the measurements of asymmetries at LEP/SLC
calls for a revision of the analysis presently used to extract physical
parameters from them in order to asset the actual meaning and limitations
of the language used (real effective $Z$ vector and axial couplings).
Typically, this language, was established for the
analysis of the asymmetries some years ago when the accuracy in the
measurements was much lower than at present and when nobody could expect such
a fast improvement of the experimental errors. One possible motivation
for this detailed analysis is
the observed discrepancy between different asymmetry measurements (namely
LEP and SLC) which although might have just a statistical origin,
could also be partly due to theoretical limitations in the language used to
express the measurements and to compare them.

    The use of the "improved Born" language to express the precision
electroweak $e^+ e^-$ data in terms of measurements of real effective
vector and axial couplings $Z$ parameters has become standard \cite{lep_ewg}
due to its conceptual
simplicity and to the fact that this language provides an easy way
of comparing and combining data from different experimental measurements.
Some studies \cite{fitting formulae} show that the use of this language
in the lineshape analysis provides an accuracy better that the one the
experimental analysis may require. Concerning the analysis of the
different asymmetries, the situation deserves still some clarification,
specially now that the accuracy for some of the asymmetries measured at
LEP/SLC reaches the permile level.
It is well known, for instance, that for the lepton
forward-backward asymmetry, the "naif" use of an "improved Born"
approximation
biases the results by an amount which is comparable to the present
experimental
accuracy. The largest piece causing this bias was already identified
some time ago \cite{yellow} to be the effect of the imaginary part
of the photon vacuum polarization but, since the experimental accuracy has
kept growing, it is justified to look to all the rest of "small" corrections
(imaginary parts of effective couplings, mass effects and heavy boxes) to
quantify how important is their effect in the parameters extracted from the
data.

    All these corrections exist since long time ago in
the literature (see for instance \cite{hollik}) and their consideration,
does not introduce any conceptual problem. The only exception is
the imaginary parts of the vector and
axial effective couplings which although well known, require some
conceptual clarification in the definition of the quoted effective couplings.

    In particular, we study in this work what are the approximations used
to define a "fitting formula" for the different asymmetries measured at
LEP/SLC. We will show that the effect of the imaginary components
of vector and axial couplings
in the frame of an effective coupling language cannot be neglected,
and we will try to modify the definition of the different
"fitting formulae" in order to take into account the relevant
effects coming from the fact that the actual effective couplings should be
complex numbers.

    The outline of this paper is as follows: first
we will review in section 2 the experimental procedure used so far
to analyze the asymmetries, and we will quantify the size of
the approximations used in the extraction of effective parameters from them.
In section 3 we will show how to improve in an easy way the simple formulae
used at present for the analysis of the asymmetries to match
the expected future experimental accuracy. In section 4 we will show the
importance of the precise definition of the effective parameters in what
concerns the handling of imaginary parts. Finally, in section 5 we will
discuss the conclusions of this analysis.

%-----------------------------------------------------------------------
\section{Analysis of the different asymmetries}
%-----------------------------------------------------------------------

In this section we will explain how the different asymmetries are
analyzed to extract from them the physical information, and what are
the approximations made in this process.

%-----------------------------------------------------------------------
\subsection{Analysis of lepton angular distribution}
%-----------------------------------------------------------------------

   The information contained in the angular distribution of
$e^+ e^- \rightarrow l^+ l^-$ with $l \neq e$ is extracted in two steps:

\begin{itemize}
\item{} first the forward-backward asymmetry is extracted from the data,
by computing directly

\begin{equation}
 \label{AFB_def}
 A_{FB} = \frac{\sigma_{F} - \sigma_{B}}{\sigma_{F} + \sigma_{B}}
\end{equation}

or by fitting the angular distribution with a simple formula

\begin{equation}
 \label{ang_dist}
 \frac{d \sigma}{d \cos\theta} = C (1+\cos^2 \theta +\frac{8}{3} A_{FB}
   \cos \theta)
\end{equation}

where, $\theta$ is taken as the polar angle in the reduced center-of-mass
frame.
The first definition has the advantage of being completely
general and, hence, applicable to any process. The second definition,
assumes a given behaviour of the
angular distribution but, under this assumption, allows a more accurate
determination of $A_{FB}$ since the whole angular distribution is used
instead of just two angular bins (as is the case in the first method).
%This results in an increase of accuracy of the order of XXXXX.
In addition, in the second procedure, if a log-likelihood
method is applied to fit the data, then there is no need of knowing
at all the angular response of the apparatus (provided it is
charge-symmetric or forward-backward symmetric).

\item{} second, the data for $A_{FB}(s)$ is fitted with a model-independent
formula in which photonic corrections are also explicitly incorporated.
This allows disentangling the pure photonic effects (the ones due to radiation
as well as the ones due to the $\gamma$ exchange contribution) from the
rest. It enables also taking properly into account the energy dependence
of the asymmetry to add-up the information from all the energy points.
The fit is done simultaneously with the lineshape data or alternatively,
by incorporating as constraints the information obtained from the lineshape
analysis. Both methods are basically equivalent provided the full covariance
matrix of lineshape parameters is used.
In this way, information about $g_V$ and $g_A$ is obtained.
These couplings can be interpreted as absorbing all sort of corrections
not explicitly accounted for in the fitting formulae, namely
real parts of $Z$ vacuum polarization, $\gamma-Z$ mixing, weak vertices
and other small effects like weak boxes .

    The imaginary part of the
$\gamma$ vacuum polarization, which plays a relevant role in $A_{FB}$, is
included explicitly in the fitting formulae. This is so since, like for the
real part, its value can be precisely predicted in QED and therefore
can be included in the photonic corrections.
On the other hand,
only the real components of effective couplings are considered in the fitting
formulae.

\end{itemize}

In this procedure, regardless on the way used to compute $A_{FB}$ from the
data, the basic assumption is that all the information of the angular
distribution is just the one in $A_{FB}$.

%The reason for this assumption being very precise for s-driven processes
%is exactly the same
%as the one allowing the use of eq.~\ref{ang_dist} to obtain it from the
%data: the fact that a formula with just one free parameter (besides the
%global normalization, which can be identified as the total cross section
%measurement) is enough two describe the angular distribution. This is due to
%the particular structure of the angular distribution in the s-channel.
At the
Born level, the angular distribution is well known that it can be written as

%\begin{eqnarray}
%\frac{d\sigma^0}{d\cos\theta}& = &
%\left(\frac{d\sigma^0}{d\cos\theta}\right)_{Z_sZ_s} +
%\left(\frac{d\sigma^0}{d\cos\theta}\right)_{Z_s\gamma_s} +
%\left(\frac{d\sigma^0}{d\cos\theta}\right)_{\gamma_s\gamma_s}
%  \nonumber \\
% & = & \frac{\pi\alpha^2}{2s} \,
%\frac{s^2}{|Z(s)|^2}
% \left[ (1+\cos^2\theta)(v^2+a^2)^2+ 8\cos\theta v^2a^2 \right]
%  \nonumber \\
% & + & \frac{\pi\alpha^2}{2s} \,
% \frac{s(s-M_Z^2)}{|Z(s)|^2}
% \left[ (1+\cos^2\theta) \, 2 \, v^2 +  \cos\theta  \, 4 \, a^2 \right]
%  \nonumber \\
% & + & \frac{\pi\alpha^2}{2s} \, (1+\cos^2\theta) \\
%       \\
%|Z(s)|^2 & = & (s-M_Z^2)^2+(M_Z\Gamma_Z)^2
%\end{eqnarray}

%so that it can be written as

\begin{equation}
\frac{d\sigma^0}{d\cos\theta}
= C^0(s) (1+\cos^2\theta +\frac{8}{3} A_{FB}(s) \cos \theta)
\end{equation}

being

\begin{equation}
C^0(s) = \frac{3}{8} \sigma^0(s) =
 \frac{\pi\alpha^2}{2s} \left[
\frac{s^2}{|Z(s)|^2} \, (v^2+a^2)^2
 + \frac{s(s-M_Z^2)}{|Z(s)|^2} \, 2 \, v^2
 + 1 \right]
\end{equation}

and

\begin{eqnarray}
 A_{FB}(s) & = & \frac{\sigma^0_{F}(s) - \sigma^0_{B}(s)}
{\sigma^0(s)}
\nonumber \\
 & = & \frac{1}{\sigma^0(s)}
 \frac{\pi\alpha^2}{2s} \left[ \frac{s^2}{|Z(s)|^2} \, 8 \, v^2 a^2
 + \frac{s(s-M_Z^2)}{|Z(s)|^2} \, 4 \, a^2 \right]
\end{eqnarray}

The picture does not change at all when the leading
non-photonic corrections are applied,
since they can be implemented by doing just some simple replacements
in the effective coupling language (see appendix), namely:

\begin{eqnarray}
\alpha &\rightarrow & \alpha(s)
   \mbox{ Real part of $\gamma$ self-energy} \nonumber \\
s(s-M_Z^2) &\rightarrow & s(s-M_Z^2)+ s^2 \frac{\Gamma_Z}{M_Z}
\Im(\Delta\alpha)
   \mbox{ Imaginary part of $\gamma$ self-energy} \nonumber \\
|Z(s)|^2 &\rightarrow  & (s-M_Z^2)^2+(s \frac{\Gamma_Z}{M_Z})^2
   \mbox{ Imaginary part of $Z$ self energy} \nonumber \\
v,a &\rightarrow & \sqrt{F_G(s)}g_V,\sqrt{F_G(s)}g_A
   \mbox{ Real part of $Z$ self energy, $\gamma-Z$ mixing, weak vertices}
    \nonumber \\
\end{eqnarray}

where
\begin{equation}
 \Im(\Delta\alpha) = \frac{\Im(\Pi^{\gamma}(s))}{1 + \Re(\Pi^{\gamma}(s))}
\end{equation}
\vspace{0.2 cm}
\begin{equation}
 F_G(s) = \frac{G_F M_Z^2}{2\sqrt{2}\pi\alpha(s)}
\end{equation}

 In such a way that, for arbitrary final state fermions
other than electrons,
we can write a fitting formula (ref.\cite{ALEPH-EW}) as:

\begin{eqnarray}
\label{FIT_FB}
\sigma^0(s) = \frac{4}{3} \ \frac{\pi\alpha^2(s)}{s} \, \, \,
( \, \, \,  Q_e^2 Q_f^2 &+& 2 \, \, Q_e Q_f g_{V_e} g_{V_f} F_G
 \frac{s(s-M_Z^2)}{ (s-M_Z^2)^2 + s^2 \frac{\Gamma_Z^2}{M_Z^2} } +
 \nonumber \\
 ((g_{V_e})^2 + (g_{A_e})^2)&\cdot&((g_{V_f})^2 + (g_{A_f})^2) \, \, F_G^2
\frac{s^2}{ (s-M_Z^2)^2 + s^2 \frac{\Gamma_Z^2}{M_Z^2} } \, \, \, )
 \nonumber \\
 \nonumber \\
A_{FB}(s)
  =  \frac{1}{\sigma^0(s)} \frac{\pi\alpha^2(s)}{s} \, \, \,
  ( \, \, \, & 2 & Q_e Q_f g_{A_e} g_{A_f} F_G
 \frac{ s(s-M_Z^2) + s\frac{\Gamma_Z}{M_Z} \Im{(\Delta\alpha}) }
 {(s-M_Z^2)^2 + s^2 \frac{\Gamma_Z^2}{M_Z^2}}
\nonumber \\
   & 4 &g_{V_e} g_{A_e} g_{V_f} g_{A_f} F_G^2
 \frac{s^2}{(s-M_Z^2)^2 + s^2 \frac{\Gamma_Z^2}{M_Z^2}} \, \, \, )
\end{eqnarray}

where the couplings $g_{V_f}$ and $g_{A_f}$ can also be re-written as:

\begin{eqnarray}
g_{V_f} & = & \rho_f^{\frac{1}{2}} \ ( I^f_3 - 2 Q_f (\sin^2\theta_W)_f )
  \nonumber \\
g_{A_f} & = & \rho_f^{\frac{1}{2}} \ I^f_3
\end{eqnarray}

$\rho_f$ and $(\sin^2\theta_W)_f$ being flavour-dependent effective
rho parameter and weak mixing angle respectively.

  To write down expression \ref{FIT_FB} some approximations have been done
\begin{itemize}

\item{} Only real parts of effective couplings are considered.
\item{} $\alpha(s)$ is considered to be real (see appendix), i.e.,
the imaginary part
of the photon self energy is neglected in the denominator
of $\alpha(s)$, and the relevant influence on the asymmetry
is explicitly taken into account
in the fitting formulae.
\item{} Corrections to $\gamma f f $ couplings are neglected.
\item{} The effective couplings are considered to be independent of s
in the fitting process.
\end{itemize}

Concerning photonic corrections, since the dominant ones are
initial state radiation,
it is rather easy to show that they do not affect sizably the simple angular
distribution described in eq.~\ref{ang_dist}. Lets take just initial state
radiation and lets include it by convoluting
the hard reduced differential
cross section with the $O(\alpha^2)$ radiator function used in the
lineshape analysis \cite{BURGERS} $H(s,x)$, where $x$ denotes the
fractional energy carried out by photon radiation:

\begin{eqnarray}
 \label{convolution}
\frac{d\sigma^0}{d\cos\theta}(s) & = &
 \int_0^{x_{max}} dx H(s,x) \left[ C^0(s(1-x)) (1+\cos^2\theta +\frac{8}{3}
 A_{FB}(s(1-x)) \cos \theta) \right] \nonumber \\ & = &
C(s) (1+\cos^2\theta +\frac{8}{3} A_{FB}(s) \cos \theta)
\end{eqnarray}

being then

\begin{eqnarray}
 \label{AFB_corrected}
C(s) & = & \int_0^{x_{max}} dx H(s,x) C^0(s(1-x)) = \frac{3}{8} \sigma(s)
\nonumber \\
A_{FB}(s) & = & \frac{1}{C(s)}
\int_0^{x_{max}} dx H(s,x) C^0(s(1-x)) A_{FB}(s(1-x))
\end{eqnarray}

This way of introducing the initial state radiation corrections, in spite of
being just a reasonable approximation in the case of the differential cross
section (eq.~\ref{convolution}),
 turns out to be a very accurate approximation
for the forward-backward asymmetry $A_{FB}$ if no strong detection cuts
are applied in the final state \cite{yellow,JADACH} and the scattering
angle used to define the forward and backward hemispheres is the one in the
centre-of-mass of the hard process. If different choices are taken
of the scattering polar angle taken to define the forward and backward regions,
then the precise form of the radiator function to be used in the
$A_{FB}$ has to be modified \cite{JADACH}.

%Therefore, the experimental analysis procedure is very clear:
%the forward-backward asymmetry is obtained for every energy by either one
%of the methods suggested and afterwards it is fitted with the radiative
%equation \ref{AFB_corrected}.

%-----------------------------------------------------------------------
\subsection{Analysis of polarized asymmetries}
%-----------------------------------------------------------------------

    The experimental procedure used to extract the polarized asymmetries from
the data is not so simple as the one used in the lepton forward-backward
asymmetry and its discussion escapes from the scope of this paper. At any rate,
when these asymmetries are already extracted from the data, then similar
arguments as the ones explained above can be used to write
fitting formulae to allow the extraction of effective parameters.
For Left-Right asymmetry defined as

\begin{equation}
\label{LR_def}
 A_{LR} = \frac{\sigma_{L} - \sigma_{R}}{\sigma_{L} + \sigma_{R}}
\end{equation}

In this case we can write

\begin{eqnarray}
\label{FIT_LR}
A_{LR}(s)
  = \frac{4}{3} \frac{1}{\sigma^0(s)} \frac{\pi\alpha^2(s)}{s} \, \, \,
  ( \, \, \, & 2 & Q_e Q_f g_{V_f} g_{A_e} F_G
 \frac{ s(s-M_Z^2) + s\frac{\Gamma_Z}{M_Z} \Im{(\Delta\alpha}) }
 {(s-M_Z^2)^2 + s^2 \frac{\Gamma_Z^2}{M_Z^2}}
\nonumber \\
   & 2 & ((g_{V_f})^2+(g_{A_f})^2) g_{V_e} g_{A_e} F_G^2
 \frac{s^2}{(s-M_Z^2)^2 + s^2 \frac{\Gamma_Z^2}{M_Z^2}} \, \, \, )
\end{eqnarray}

    This formula has exactly the same approximations used to
write \ref{FIT_FB}. In the same way, one can write a fitting formula
for the polarized
Forward-Backward asymmetry defined as

\begin{equation}
 \label{FBLR_def}
 A_{LR}^{FB} = \frac{(\sigma_{L} - \sigma_{R})_{F}-
(\sigma_{L} - \sigma_{R})_{B}}{\sigma^0(s)}
\end{equation}

by doing the replacements in formula \ref{FIT_LR}

\begin{eqnarray}
\nonumber g_{V_e} &\rightarrow & g_{V_f} \nonumber \\
\nonumber g_{A_e} &\rightarrow & g_{A_f} \nonumber \\
\nonumber g_{V_f} &\rightarrow & g_{V_e} \nonumber \\
\nonumber g_{A_f} &\rightarrow & g_{A_e} \nonumber \\
\end{eqnarray}

i.e. interchanging the roles between initial and final fermions, and removing
the factor $\frac{4}{3}$.

\begin{eqnarray}
\label{FIT_FBLR}
A_{LR}^{FB}(s)
  =  \frac{1}{\sigma^0(s)} \frac{\pi\alpha^2(s)}{s} \, \, \,
  ( \, \, \, & 2 & Q_f Q_e g_{V_e} g_{A_f} F_G
 \frac{ s(s-M_Z^2) + s\frac{\Gamma_Z}{M_Z} \Im{(\Delta\alpha}) }
 {(s-M_Z^2)^2 + s^2 \frac{\Gamma_Z^2}{M_Z^2}}
\nonumber \\
   & 2 &((g_{V_e})^2+(g_{A_e})^2) g_{V_f} g_{A_f} F_G^2
 \frac{s^2}{(s-M_Z^2)^2 + s^2 \frac{\Gamma_Z^2}{M_Z^2}} \, \, \, )
\end{eqnarray}

%-----------------------------------------------------------------------
\subsection{Accuracy of the fitting formulae compared with SM
predictions.}
%-----------------------------------------------------------------------

    In this section we will compute the actual accuracy of the above
expressions \ref{FIT_FB},\ref{FIT_LR},\ref{FIT_FBLR},
by comparing their results with the ones obtained
using a complete electroweak library (in our case BHM \cite{BHM}).
One can write a complete expression for the different asymmetries
defined above, neglecting mass terms and box diagram contributions as
\cite{hollik}:

\begin{equation}
 A_{FB} = \frac{3}{4} \frac{G_{3}(s)}{G_{1}(s)}
\end{equation}
\vspace{0.3cm}
\begin{equation}
 A_{LR} = \frac{H_{1}(s)}{G_{1}(s)}
\end{equation}
\vspace{0.3cm}
\begin{equation}
 A_{LR}^{FB} = \frac{3}{4} \frac{H_{3}(s)}{G_{1}(s)}
\end{equation}

where

\begin{equation}
 G_1 = \Re {\sum_{j,k=1}^{2} {( V_j^e   V_k^{e*}  +  A_j^e   A_k^{e*} )
( V_j^f   V_k^{f*}  +  A_j^f   A_k^{f*} ) \chi _j \chi _k^{*}}}
\end{equation}
\vspace{0.3cm}
\begin{equation}
 G_3 = \Re{\sum_{j,k=1}^{2} {( V_j^e   A_k^{e*}  +  A_j^e   V_k^{e*} )
( V_j^f   A_k^{f*}  +  A_j^f   V_k^{f*} ) \chi _j \chi _k^{*}}}
\end{equation}
\vspace{0.3cm}
\begin{equation}
 H_1 = \Re{\sum_{j,k=1}^{2} {( V_j^e   A_k^{e*}  +  A_j^e   V_k^{e*} )
( V_j^f   V_k^{f*}  +  A_j^f   A_k^{f*} ) \chi_j \chi_k^{*}}}
\end{equation}
\vspace{0.3cm}
\begin{equation}
 H_3 = \Re{\sum_{j,k=1}^{2} {( V_j^e   V_k^{e*}  +  A_j^e   A_k^{e*} )
( V_j^f   A_k^{f*}  +  A_j^f   V_k^{f*} ) \chi_j \chi_k^{*}}}
\end{equation}
\vspace{0.3cm}

according to the following table:

\begin{table}[htb]
%\vspace{2cm}
\begin{center}
\begin{tabular}{|l|c|c|c|c|r|}
\hline
{\em j} & {\em $V_j^e$} & {\em $A_j^e$} & {\em $V_j^f$} & {\em $A_j^f$}
& {\em $\chi_j$}  \\
\hline
1 & ($Q^{e}$ + $F_{V \gamma e}$) & -$F_{A \gamma e}$
& ($Q^{f}$ + $F_{V \gamma f}$) & -$F_{A \gamma f}$  & $\chi_1$ \\
2 & $g_{V_e}$ & $g_{A_e}$ & $g_{V_f}$ & $g_{A_f}$ & $\chi_2$ \\
\hline
\end{tabular}
\end{center}
\caption[]
{\protect\footnotesize}
%\vspace{2cm}
\end{table}

and

\begin{equation}
 \chi_1 = \frac{4 \pi \alpha}{(1+\Re{\Pi^\gamma})+i\Im{\Pi^\gamma}}
\end{equation}
\vspace{0.3cm}
\begin{equation}
 \chi_2 = \frac{\sqrt{2} G_F M_Z^2 s}{(s - M_Z^2)+i s \frac{\Gamma_Z}{M_Z}}
\end{equation}
\vspace{0.3cm}

 By computing $G_i$ and $H_i$ in this way, we have made the choice of
factorizing the initial and final state weak vertex corrections. This allows to
preserve the Born structure of the calculation in the sense that only two
``dressed'' amplitudes are considered: the photon exchange and the Z exchange
one. This choice is different from the one in reference \cite{hollik} in which
the calculation (forgetting weak boxes) involves as much as 8 amplitudes. The
numerical differences of both approaches referring to asymmetries predictions
are of the order of $10^{-4}$, and this is of the same level than the
theoretical uncertainties from different treatments of higher order  radiative
corrections.

We have included\footnote{All the numerical evaluations is this paper have been
done for $M_z$ = 91.187 GeV, $\alpha_s$ = 0.12, $M_t$ = 150 GeV and $M_h$ = 300
GeV and NOT taking into account QED and final state QCD corrections} also in
the calculation of these expressions the contributions due to heavy box
diagrams and mass terms following reference \cite{hollik}. The effect due to
these two sources of corrections can be seen in table 2 in which one can also
see the effect of switching off completely the imaginary part of the vector and
axial couplings of table 1.

\begin{table}[htb]
%\vspace{2cm}
\begin{center}
\begin{tabular}{|l|c|c|c|c|c|}
\hline
   & Complete & No Box. & No Mass &  No Imag.
& Exper. acc. \\
\hline
 & & $\times 10^{-5}$ & $\times 10^{-5}$ & $\times 10^{-5}$
& $\times 10^{-5}$ \\
\hline \hline
$A_{FB}(M_z^2)$ & 0.01534  & $ \leq $ 1 & -1 & +42 & 160 \\
\hline
$A_{LR}(M_z^2)$ & 0.13348  & $ \leq $ 1 & -20 & +187 & 600 \\
\hline
$A_{LR}^{FB}(M_z^2)$ & 0.10004 & -1 & -8 & +139 & 1200 \\
\hline
\end{tabular}
\end{center}
\caption[]
{\protect\footnotesize Effect of switching off different ``small corrections'':
heavy boxes, mass terms and imaginary parts of effective couplings.
The mass effects are quoted for the tau lepton.
Present experimental errors are quoted in the last column as a reference
(ref.\cite{EW_paper}).}
%\vspace{2cm}
\end{table}

    In table 3 we compare the complete result (including boxes, mass
corrections and all imaginary parts) with
the results using the simple formulae \ref{FIT_FB},\ref{FIT_LR},
\ref{FIT_FBLR}. The differences ($\Delta$) are shown
for leptons, c-quarks, s-quarks and b-quarks. Also some
experimental errors are quoted in parenthesis for some asymmetries
(ref.\cite{EW_paper}) to have a reference for the importance of the
differences. The dependence of these differences with energy can be seen
in fig.1 for leptons.

\begin{table}[htb]
%\vspace{2cm}
\begin{center}
\begin{tabular}{|l|c|c|c|c|}
\hline
  & {\em $\Delta_{leptons}$} & {\em $\Delta_{c-quarks}$} &
{\em $\Delta_{s-quarks}$} & {\em $\Delta_{b-quarks}$}   \\
\hline
$A_{FB}(M_z^2)$ & -0.00043 (0.0016) & -0.00023 (0.011)& +0.00000
& +0.00006 (0.0043)\\
\hline
$A_{LR}(M_z^2)$ & -0.00165 (0.006) & -0.00058 & +0.00009 & +0.00096\\
\hline
$A_{LR}^{FB}(M_z^2)$ & -0.00131 (0.012) & -0.00051 & -0.00010 & -0.00002\\
\hline
\end{tabular}
\end{center}
\caption[]
{\protect\footnotesize Differences between fitting formulae with real
effective couplings and SM predictions for the three asymmetries. Experimental
errors are quoted in parenthesis for the asymmetries with existing experimental
determinations (ref.\cite{EW_paper}).}
%\vspace{2cm}
\end{table}

What we can learn from
the results in table 3 is that at least for $A_{FB}^l$ and $A_{LR}^l$ the
difference is not negligible compared with the present experimental error.
So at this point it's apparent the necessity to improve expressions
\ref{FIT_FB},\ref{FIT_LR},\ref{FIT_FBLR} in order to have theoretical
uncertainties far away from experimental errors.

%-----------------------------------------------------------------------
\section{Improved analysis taking into account the relevant imaginary
components.}
%-----------------------------------------------------------------------

By far, of the approximations
listed in section 2.1, the fact that we are neglecting the
imaginary parts of the effective
couplings is the cause of differences shown in table 2.
In this section we are going to see how we can keep the simplicity of
these formulae and, at the same time
take into account the relevant imaginary components in an approximated
but very accurate way.

%-----------------------------------------------------------------------
\subsection{Modifications to fitting formulae.}
%-----------------------------------------------------------------------

    A numerical investigation shows that essentially all the difference
comes from the interference between $\gamma-Z$ channels in the numerator
of the different asymmetries. This is something we should expect, because the
interference term is the only place where the imaginary components, which are
small corrections, appear linearly (in the rest they appear quadratically).

    On the other hand, it's well known that $\Re(g_A)$ is larger than
$\Re(g_V)$ \footnote{This is true also for b quarks, although the differences
 between $\Re(g_A)$ and $\Re(g_V)$ are smaller.}. So, with this in mind it's
easy to understand that the main contributions come from imaginary terms that
multiply $\Re(g_A)$.

    In this way we can improve formulae
\ref{FIT_FB},\ref{FIT_LR},\ref{FIT_FBLR} by simply doing the replacement:

\begin{eqnarray}
\label{modif_1}
\Im(\Delta \alpha) &\rightarrow & \Im(\Delta \alpha +\Delta g)
\end{eqnarray}

where $\Im(\Delta g)$ is respectively \footnote{In the case of the
forward-backward asymmetry
we have considered also the effect coming from $\Im(F_v)$ because
it corresponds to $20\%$ of the difference quoted in
table 2. In addition, in this case the effects coming from the imaginary
components of the photon formfactors are more important than the ones
originated by the real components. This is due to the structure of the
interference terms, where
real components of $g$ are merged with imaginary components of $F$.}:

\begin{eqnarray}
\label{modif_2}
 \Im(\Delta g)_{FB} & = &
 (\frac{\Im(g_{A_e})}{\Re(g_{A_e})}+
\frac{\Im(g_{A_f})}{\Re(g_{A_f})}) + (\frac{\Im(F_{V \gamma e})}{Q_e} +
\frac{\Im(F_{V \gamma f})}{Q_f})
\nonumber \\
 \Im(\Delta g)_{LR} & = &
 \frac{\Im(g_{V_f})}{\Re(g_{V_f})}
\nonumber \\
 \Im(\Delta g)_{LR}^{FB} & = &
 \frac{\Im(g_{V_e})}{\Re(g_{V_e})}
\end{eqnarray}

    As a numerical example if one evaluates this modifications for leptons
the results are:

\begin{eqnarray}
 \Im(\Delta \alpha) & \sim & +0.017732  \nonumber \\
 \Im(\Delta g)_{FB}  \sim  -0.004667 + 0.000714 & \sim & -0.003953 \nonumber \\
 \Im(\Delta g)_{LR} \sim  \Im(\Delta g)_{LR}^{FB} &\sim& -0.181936 \nonumber \\
\end{eqnarray}

    One can see from the above numbers that the main contributions come
from  $\Im(g_{A_f})$ and  $\Im(g_{V_f})$.
Following the definitions in the appendix, one can see that $\Im(g_{A_f})$
comes from $\Im(F_{A Zf})$ (i.e., the formfactor of the $Z f f$ vertex
that incorporates weak vertex corrections and external fermion self energies),
and thus it is flavour dependent. Reversely, $\Im(g_{V_f})$ comes essentially
from the universal corrections to $s_W^2$, namely from $\Im(\Pi^{\gamma Z})$
(i.e. the self energy of $\gamma-Z$ mixing), but also in this case the flavour
dependent part is not negligible.

\begin{table}[htb]
%\vspace{2cm}{}
\begin{center}
\begin{tabular}{|l|c|c|c|c|}
\hline
  & {\em $\Delta_{leptons}$} & {\em $\Delta_{c-quarks}$} &
{\em $\Delta_{s-quarks}$} & {\em $\Delta_{b-quarks}$}   \\
\hline
$A_{FB}(M_z^2)$ & +0.00003 & +0.00004 & +0.00013 & +0.00015\\
\hline
$A_{LR}(M_z^2)$ & +0.00024 & +0.00018 & +0.00005 & +0.00105\\
\hline
$A_{LR}^{FB}(M_z^2)$ & +0.00011 & +0.00033 & +0.00023 & +0.00034\\
\hline
\end{tabular}
\end{center}
\caption[]
{\protect\footnotesize Differences between fitting formulae after the
modifications explained in the text and SM predictions for the three
asymmetries}
%\vspace{2cm}
\end{table}

  After the implementation of the above modifications
(\ref{modif_1},\ref{modif_2}) the level of
precision is one order of magnitude higher for the asymmetries were the
experimental accuracy is better as one can see in
table 3 and therefore copes with the expected accuracy.
We can also see in fig.1 the precision of the approximation
before and after modifications as a function of the energy in the
case of leptons.

%-----------------------------------------------------------------------
\section{Precise definition of fitting parameters}
%-----------------------------------------------------------------------

    The precision that asymmetry measurements are presently achieving requires
a more precise definition of the actual meaning of the parameters extracted
from these measurements. For instance, differences in the peak lepton
forward-backward asymmetry at the level of 0.0008
have been observed when comparing the results obtained
using two of the most popular electroweak libraries at LEP, (BHM and
ZFITTER) \cite{ZHONG}. This is about
$40\%$ of the one sigma error at LEP, and it propagates to a difference in
the effective mixing angle of $\Delta \sin^2(\theta)_{eff}^{lept} \sim 0.0004$.
    We will show in this section that, in fact, the part of this difference
which does not come from differences in the treatment of QED corrections
( $\sim$ 0.00035 ), can be explained by the different treatment
of the imaginary components of the effective couplings ( $\sim$ 0.00025 ) plus
differences in the predictions of the electroweak libraries
related to the treatment of higher orders ( $\sim$ 0.0001 ).

    From the theoretical point of view there are some conceptual
differences in the definitions of the effective couplings between
the two aforementioned libraries \cite{HOLLIK-BARDIN}.
In ZFITTER the small formfactors $F_{V \gamma f}$ and $F_{A \gamma f}$
(due to weak vertex corrections to the photon coupling) and the bosonic
loops to
the photon propagator have been implemented in the Z boson amplitude to have a
gauge invariant splitting between the photon and the $Z^0$ amplitude
\cite{ZFITTER}. The price to pay is that a vector coupling $g_{V_{e,f}}$
which depends on both initial and final state in a non-factorizable way
shows up in the Z amplitude. In BHM, these small contributions are kept
in the photon amplitude for simplicity and, in spite that, then the
splitting is not gauge invariant, numerically this does
not have any consequence.
This makes the two definitions for $g_V$ and $g_A$ ``in principle'' different.
``In practice'', the effects due to the different treatment of
$F_{V \gamma f}$ and $F_{A \gamma f}$ are numerically irrelevant:
of the order of $\Delta g_A \sim 10^{-9}$ and $\Delta g_V \sim 10^{-5}$ but
as we shall show, this fact introduces conceptual complications in
the definition of which are the actual parameters extracted
by the experiments.

    A part from this, the fact that the two libraries use different
calculationnal schemes to implement the calculations, and that there are
some differences in the treatment of higher order corrections, originates
numerically differences in $g_V$ and $g_A$ of the order of
$\Delta g \sim 10^{-4}$. This propagates to a difference in the
prediction of the leptons forward-backward asymmetries of the
order of 0.0001.

%    At the level of precision of asymmetries measurements nowadays, one has
%to remember that the effective couplings are complex numbers, and it is not
%anymore a good approximation continue neglecting their imaginary components.
%As the physical origin of these imaginary components is related to absorptive
%parts of the matrix element, it can be considered "known physics"
%because only particles under the Z mass can contribute, and therefore
%one can decide to fix this terms to the Standard Model predictions.

    We have shown in the previous section the importance of
the imaginary components of the effective couplings for some asymmetries.
Given the fact that the data is not sensitive enough to measure the
effective couplings as complex numbers, just real effective coupling are
extracted but then the common procedure followed in the above libraries is
borrowing the imaginary parts of the effective couplings from a complete
explicit Standard Model calculation.

    This procedure guarantees the same accuracy as the complete calculation
for the Standard Model interpretation of the parameters. Given the fact that
the imaginary parts are small and that, due to their absorptive origin,
they depend only mildly on the model, this procedure spoils only very
softly the ``model independence'' of the methode.

    At any rate, the practical implementation of this procedure
introduces some ambiguities in the actual meaning of the extracted
parameters since they may be defined as:
\begin{itemize}
\item{} the modulus of the complex vector and axial effective couplings,
the phase being taken from the Standard Model calculation or,
\item{} the real parts of the vector and axial effective couplings,
their imaginary parts being taken from the Standard Model calculation or even
\item{} the real parts of the vector and axial effective couplings computed
from the real part of $\rho$ and $\sin^2\theta_{eff}^{lept}$, the imaginary
parts of $\rho$ and $\sin^2\theta_{eff}^{lept}$ being taken from the Standard
Model.
\end{itemize}

    Given the small relative size of the imaginary components with respect to
the real ones, the first two approaches are numerically equivalent and, in fact
the second one is the BHM choice. In the
ZFITTER package one as also the option to fit $\Re(g_V)$ and $\Re(g_A)$
as defined in
equation 1, but in this case the imaginary components of the vector and axial
effective couplings are not fixed to the Standard Model predictions, but
instead what is fixed are the
imaginary components of the $\kappa$ and $\rho$ parameters because
these are the "natural" parameters that the program uses internally.
This, of course, gives different results but still the effect is much below
the differences we were quoting.

In fact, what produces the bulk of that difference in the lepton
forward-backward asymmetry is whether one understands
the extracted vector parameters as the real part of the vector couplings
for each flavour

             $$ g_{V_e} g_{V_f} (experimental) = \Re(g_{V_e}) \Re(g_{V_f}) $$

or rather as the real part of the initial-final state vector coupling

             $$ g_{V_e} g_{V_f} (experimental)  = \Re(g_{V_{e,f}}) $$

At $\sqrt{s}=M_Z$, in very good approximation $g_{V e,f} \cong g_{V_e} g_{V_f}$
\cite{ZFITTER} but then,

  $$ \Re(g_{V_{e,f}}) = \Re(g_{V_e})\Re(g_{V_f}) - \Im(g_{V_e})\Im(g_{V_f}) $$

so that both definitions differ when the imaginary parts are not neglected.
It turns out that this product of imaginary parts produces a
difference in the predicted lepton forward-backward asymmetry at the level of
$\sim$ 0.00025 which is what was missing to explain the difference between
BHM and ZFITTER.
So, in order to be consistent, one has to choose whether the fitting parameters
are ($\Re(g_{V_f}),\Re(g_{A_f})$) or any of the different alternative options
described above \footnote{In our opinion, using ($\Re(g_{V_f}),\Re(g_{A_f})$)
matches best the language already used by the experiments and has the simplest
connection with the Born formulae and therefore this is the definition that we
advocate}, but it is not more true that at the level of accuracy we want to
achieve all these approaches are still equivalent.

%-----------------------------------------------------------------------
\section{Conclusions}
%-----------------------------------------------------------------------

    From the present study, the main conclusion is
 that at the present level of experimental accuracy in the asymmetry
measurements, some of the approximations
used in the past to define a "fitting formula" need to be reviewed. In
particular we have seen how we can take into account the relevant effects
coming from the imaginary components of effective couplings with minimal
changes in the definition of a "fitting formula".
We have also shown
that the use of different prescriptions on how to handle
the imaginary parts when quoting just real effective couplings from the data
may produce sizable differences. A definition of the real effective couplings
specifying how to handle the imaginary parts has been advocated.

%-----------------------------------------------------------------------
\section*{Appendix: Writing the amplitudes for $e^+e^- \rightarrow f^+f^-$
in an effective coupling language.}
%-----------------------------------------------------------------------
It has been shown by several groups that in four fermion processes, the
radiatively corrected matrix element squared can be rewritten keeping a
Born-like structure by defining running effective complex parameters
\cite{lynn},\cite{HOLLIK-BARDIN},\cite{hollik}.

  This fact is specially transparent for neutral current processes in which
the non-photonic corrections at one-loop level separate
naturally from the photonic ones forming a gauge-invariant subset.

  Initial-final state factorizable corrections such as self-energies
and vertex corrections can be easily absorbed by redefining the Born couplings
as we shall see. Concerning non-factorizable corrections such as boxes,
two different approaches exist:

\begin{itemize}
\item{} Absorbing them also in the definition of the effective parameters
\cite{HOLLIK-BARDIN}. The price to pay is that some effective parameters
become not just a function of $s$ but
also of $\cos\theta$ and, in addition, the Born-like structure is somewhat
spoiled by the presence of effective parameters which do not show up
in the pure Born approach.
\item{} Keeping them out from the definition of the effective parameters
\cite{hollik}. They must be included as explicit corrections afterwards.
This approach has the advantage of being very simple and producing a set
of effective parameters which depend only on $s$ and have a clear Born
interpretation. Nevertheless, the price
to pay is that, in this case, the effective parameters are defined in a
gauge non-invariant way so that attention should be paid to
the gauge choice. \footnote{For a while, this observation prevented
the theoretical community from accepting the usefulness of the effective
parameter approach which, at the end, has been the one chosen by the
experimental community to perform the measurements.
In fact, if the predictions are computed in a gauge in which the non-absorbed
corrections are numerically irrelevant (as is the case for the 't Hooft-Feynman
gauge) the calculations using these effective parameters produce, in fact,
numerically gauge-invariant results}.
\end{itemize}

  For reasons of simplicity we will follow the second approach to define
the meaning of the effective couplings.

%-----------------------------------------------------------------------
\subsection*{Universal effective parameters.}
%-----------------------------------------------------------------------

  In the 't Hooft-Feynman gauge the
dominant corrections are by far, the vector boson vacuum polarizations. Since
these corrections do not depend of the species of the external fermions they
are, in fact, universal (process independent). The three dressed self
energies showing up in $Z$ processes, can be absorbed in three universal
parameters \cite{hollik}:

\begin{itemize}
\item{} the photon self energy $\Pi^{\gamma}(s)$ is absorbed in an effective
coupling constant $\alpha(s)$, defined as:

   $$ \bar{\alpha}(s) = \frac{\alpha_0}{1+\Pi^{\gamma}(s)} $$

It is important to stress that, since $\Pi^{\gamma}(s)$ is a complex function,
so is $\bar{\alpha}(s)$. Nevertheless, since the imaginary part of
$\bar{\alpha}(s)$
is small compared with the real one, its main effect happens in
observables which are sensitive to phase differences between photon and $Z$
exchange diagrams such as the forward-backward charge asymmetry.

\item{} the way the $\gamma Z$ mixing is treated is slightly more complicated.
Since the $\gamma Z$ mixing does not show up at tree level, to keep the Born
structure, it must be absorbed in the neutral current coupling parameters.
Then the neutral current is redefined as

   $$ C^Z_{\mu} = \gamma_{\mu} (v_f - a_f \gamma_5 ) + \gamma_{\mu} Q_f
   \frac{\Pi_{\gamma Z}(s)}{1+\Pi^{\gamma}(s)} =
     \frac{1}{2 s_W c_W} \left[ \gamma_{\mu} ( I_3^f - 2 Q_f \bar{s}^2_W(s) )
     - I_3^f \gamma_{\mu} \gamma_5 \right] $$

being

    $$ \bar{s}^2_W(s) = s_W^2 ( 1 + \Delta\kappa(s) ) $$

the complex universal effective weak mixing angle, where

 $$ \Delta\kappa(s) =
   -\frac{c_W}{s_W}\frac{\Pi^{\gamma Z}(s)}{1+\Pi^{\gamma}(s)} $$

\item{} finally, the $Z$ self energy has to be absorbed into a third parameter.
The problem is that we have already used the only two tree level parameter
($\alpha$ and $s_W$) to absorb vacuum polarization corrections. The way out is
the following: lets first consider the corrected $Z$ propagator with the
overall
coupling constant that will come from the initial and final state particle
couplings

 $$  \frac{e^2}{4 s_W^2 c_W^2}
     \frac{1}{s - M_Z^2 + \Re(\Sigma_Z(s)) + i \Im(\Sigma_Z(s))} =
  \frac{e^2}{4 s_W^2 c_W^2} \frac{1}{1+\Pi^Z(s)}
     \frac{1}{s - M_Z^2 + i \frac{\Im(\Sigma_Z(s))}{1+\Pi^Z(s)}} $$

where

$$ \Pi^Z(s) = \frac{\Re(\Sigma_Z(s))}{s - M_Z^2} $$

Therefore, the factor  $\frac{e^2}{4 s_W^2 c_W^2} \frac{1}{1+\Pi^Z(s)}$
which multiplies the Breit-Wigner like propagator,
can be considered as the effective strength of the purely weak interactions
and an appropriate way of introducing a running parameter to account for it is
recalling the tree level $G_F$ relation:

 $$ \frac{e^2}{4 s_W^2 c_W^2} = \sqrt{2} G_F M_Z^2 (1 - \Delta r) \rho_0$$

where $\rho_0$ is the tree level $\rho$ parameter which in the Minimal Standard
Model is exactly $1$. Therefore we can write

$$ \frac{e^2}{4 s_W^2 c_W^2} \frac{1}{1+\Pi^Z(s)} = \sqrt{2} G_F M_Z^2
\rho(s)$$

being

$$ \bar{\rho}(s) = \rho_0 (1 + \Delta\rho(s)) $$

the universal effective $\rho$ parameter, where

$$ \Delta\rho(s) = \frac{1-\Delta r}{1+\Pi^Z(s)} - 1 $$

$\Delta\rho(s)$ and hence $\bar{\rho}(s)$ are real quantities by definition,
since the imaginary part of the $Z$ vacuum polarization will be treated
separately.
It is important to stress that this $\Delta\rho$ is numerically and
conceptually
different from the one introduced when discussing $\Delta r$. The one
introduced
there accounted for the ratio of $W$ to $Z$ vacuum polarization corrections at
$q^2 \cong 0$, whereas the one introduced now is more complex since accounts
for the ratio of the whole $\Delta r$ correction (which in spite that is
dominated by the $W$ vacuum polarization at $q^2 \cong 0$, includes also
sizable QED corrections ) to the real part of the derivative of the
$Z$ vacuum polarization at $q^2 = s$.
Nevertheless, the coefficient of the dominant $m_t$ terms is the same.

Finally, it can be shown \cite{hollik} that the imaginary part of the $Z$
self energy can be interpreted, through the use of the optical theorem as

 $$\frac{\Im(\Sigma_Z(s))}{1+\Pi^Z(s)}
   = \bar{\Gamma}_Z(s) \cong \frac{s}{M_Z} \bar{\Gamma}_Z(M_Z^2)$$

where $\bar{\Gamma}_Z(s)$ stands for the Born total $Z$ decay width in
terms of effective couplings.

\end{itemize}

Summarizing, for $Z$ physics, the vacuum polarization corrections
can be properly included by writing the Born amplitudes in terms
of the effective running couplings $\bar{\alpha(s)}$, $\bar{\rho(s)}$ and
$\bar{s}_W^2(s)$ as:

\begin{eqnarray}
  \bar{A} &=& Q_e Q_f \frac{4\pi\bar{\alpha}(s)}{s}
   [ \gamma_{\mu} \otimes \gamma^{\mu} ]
    \, \,  +  \, \, \sqrt{2} G_F M_Z^2 \bar{\rho}(s)
  \frac{1}{s-M_Z^2+i s \frac{\bar{\Gamma}_Z}{M_Z}}  \nonumber \\
   && [ \gamma_{\mu}(I_3^e - 2 Q_e \bar{s}_W^2(s))-\gamma_{\mu}\gamma_5 I_3^e ]
  \otimes
   [ \gamma^{\mu}(I_3^f - 2 Q_f \bar{s}_W^2(s))-\gamma^{\mu}\gamma^5 I_3^f ]
   \nonumber
\end{eqnarray}

This representation of the amplitudes is accurate for the calculation of any
observable \footnote{Exception made of the observables for $b$ quarks for
which, as pointed out before, vertex corrections play an important role}
at the percent level. It is worth mentioning also that at this level, the
definition of the effective amplitudes in the
most popular electroweak libraries \cite{BHM,ZFITTER} conceptually agree.

%-----------------------------------------------------------------------
\subsection*{Flavour-dependent effective parameters.}
%-----------------------------------------------------------------------

 The fact that the accuracy for some $Z$ observables reaches the permile level
requires the consideration of the next level of corrections, namely the weak
vertex ones. These corrections, unlike the vacuum polarization ones, depend
explicitly on the species of the external fermions and therefore are flavour
dependent. Hence, they can be absorbed into effective parameters at the price
of making them flavour dependent, that is, having a set of effective parameters
for every fermion species.

 As of the photon exchange amplitude, the current including vertex
corrections at one loop can be written as:

 $$  C^{\gamma}_{\mu} = \left[ \gamma_{\mu}
    \left( Q^f + F_{V \gamma f}(s) - F_{A \gamma f}(s)  \gamma_5 \right)
         \right] $$

where $F_{V \gamma f}$ and $F_{A \gamma f}$ are the complex vector and axial
photon formfactors (see for instance \cite{hollik}).

 In the case of the weak vertex corrections for the $Z$ amplitudes, at one
loop we can introduce them through the use of complex form factors
modifying the Born axial and vector couplings in the currents
\cite{hollik}:

 $$  C^Z_{\mu} = \left[ \gamma_{\mu}
    \left( v_f + F_{V Zf}(s) - (a_f + F_{A Zf}(s))  \gamma_5 \right)
 + \gamma_{\mu} Q_f
   \frac{\Pi_{\gamma Z}(s)}{1+\Pi^{\gamma}(s)} \right] $$

where $F_{V \gamma Z}$ and $F_{A \gamma Z}$ are the complex vector and axial
$Z$ formfactors and
the $\gamma Z$ mixing correction has also been included explicitly. This
can be rewritten as

    $$ \left( 1 + \frac{F_{A Zf}(s)}{a_f} \right)
      \frac{1}{2 s_W c_W} [ \gamma_{\mu} ( I_3^f - 2 Q_f s^2_{Wf}(s) )
     - I_3^f \gamma_{\mu} \gamma_5 ] $$

being

    $$ s^2_{Wf}(s) = s_W^2 ( 1 + \Delta\kappa(s) + \Delta\kappa_f(s) ) $$

the flavour-dependent complex effective weak mixing angle, where

 $$ \Delta\kappa_f(s) =
   -\frac{1}{Q_f}\frac{c_W}{s_W}
       \left( F_{V Zf}(s) - \frac{v_f}{a_f} F_{A Zf}(s) \right) $$

is the flavour-dependent vertex correction.

  After this algebra, the effective strength of the purely weak interactions
becomes

 $$ \frac{e^2}{4 s_W^2 c_W^2} \frac{1}{1+\Pi^Z(s)}
    \left( 1 + \frac{F_{A Ze}(s)}{a_e} \right)
    \left( 1 + \frac{F_{A Zf}(s)}{a_f} \right) = \sqrt{2} G_F M_Z^2 (\rho_e(s)
    \rho_f(s))^{\frac{1}{2}} $$

being

$$ \rho_f(s) = \rho_0 (1 + \Delta\rho(s) + \Delta\rho_f(s)) $$

the flavour-dependent complex effective $\rho$ parameter, where

$$ \Delta\rho_f(s) = \left( 1 + \frac{F_{A Zf}(s)}{a_f} \right)^2 - 1 $$

is the flavour-dependent complex vertex correction.

With the introduction of these complex flavour-dependent
effective parameters, the $Z$ exchange amplitude can simply be written as:

     $$ A_Z = \sqrt{2} G_F M_Z^2
  \frac{1}{s-M_Z^2+i s \frac{\Gamma_Z}{M_Z}}
   [ \gamma_{\mu}(g_{V_e}(s) - g_{A_e}(s)\gamma_5) ]
  \otimes
   [ \gamma^{\mu}(g_{V_f}(s) - g_{A_f}(s)\gamma^5) ] $$

where the complex effective vector and axial couplings are defined as

\begin{eqnarray}
  g_{V_f}(s) &=& \sqrt{\rho_f} ( I_3^f - 2 Q_f s_{Wf}^2 ) \nonumber \\
  g_{A_f}(s) &=& \sqrt{\rho_f} I_3^f \nonumber
\end{eqnarray}

%-----------------------------------------------------------------------
\section*{Acknowledgments}
%-----------------------------------------------------------------------

 We want to thank W.Hollik and D.Bardin for fruitful discussions about
this topic. We are also indebted to W.Chen and Z.Feng for their early
help in the comparison between BHM and ZFITTER.

%%%%%%%%%%%%%%%%%%%%%%%%%%%%%%%%%%%%%%%%%%%%%%%%%%%%%%%%%%%%%%%%%%%%%%%%%%%%
\begin{figure}[htb]
\begin{center}
\mbox{
\epsfig{file=fig1.eps,height=18cm}}
\end{center}
\caption[]
{\protect\footnotesize
Absolute value of differences between the SM predictions for the different
asymmetries and the
"fitting formulae" before and after modifications explained in the text.}
\label{afb_fig}
\end{figure}
%%%%%%%%%%%%%%%%%%%%%%%%%%%%%%%%%%%%%%%%%%%%%%%%%%%%%%%%%%%%%%%%%%%%%%%%%%%%

%-----------------------------------------------------------------------

%-----------------------------------------------------------------------

\clearpage
\newpage
\begin{thebibliography}{99}
    \bibitem{lep_ewg}
            The LEP collaborations: ALEPH, DELPHI, L3 and OPAL,
            Phys. Lett. {\bf B276} (1992) 247
    \bibitem{fitting formulae}
            M.Mart\'\i nez et al., Z. Phys. C - Particles and Fields
            {\bf 49} (1991) 645;\\
            S.Jadach et al.,  Phys. Lett. {\bf B280} (1992) 129;\\
            M.Mart\'\i nez and B.Pietrzyk, Phys. Lett. {\bf B324} (1994) 492.
    \bibitem{yellow}
            M.B\"ohm , W. Hollik et al., Forward-Backward asymmetries,
            Proceedings of the Workshop on Z Physics
            at LEP1, CERN 89-08, Sept. 1989, ed. G. Altarelli et al.,Vol. 1 ,
            p. 203.
    \bibitem{hollik}
            W. Hollik , Radiative Corrections in the Standard Model and their
            role for precision tests of the
            Electroweak Theory, Fortschr. Phys.  {\bf 38} (1990) 165.
    \bibitem{EW_paper}
            The LEP Electroweak Working Group, Internal Note summarizing the
            combination of preliminary LEP data for
            La Thuile and Moriond conferences, March 1994.
    \bibitem{ALEPH-EW}
            ALEPH collaboration, Z. Phys. C - Particles and Fields  {\bf 53}
            (1992) 1
    \bibitem{BURGERS}
            F.A.Berends, W.L.Van Neerven and G.J.H.Burgers,
            Nucl.Phys.{\bf B297} (1988) 429; {\bf B304} (1988) 921 (E)
    \bibitem{JADACH}
            S. Jadach, Z. Was et al., The tau polarization measurement,
            Proceedings of the Workshop on Z Physics
            at LEP1, CERN 89-08, Sept. 1989, ed. G. Altarelli et al.,Vol. 1 ,
            p. 235.
    \bibitem{BHM}
            Computer code by G. Burgers, W. Hollik and M. Mart\'\i nez;\\
            M. Consoli, W. Hollik and F. Jegerlehner:
            Proceedings of the Workshop on Z Physics at LEP1, CERN 89-08,
            Sept. 1989, ed. G. Altarelli et al.,Vol. 1 , p. 7.\\
            G. Burgers, F. Jegerlehner, B. Kniehl and J. K\"uhn :
            the same proceedings Vol. 1 , p. 55.
    \bibitem{ZHONG} W.Chen and Z.Feng, private communication.
    \bibitem{HOLLIK-BARDIN}
            D. Bardin, W. Hollik and T. Riemann, Bhabha scattering with
            higher order weak loop corrections,
            Z. Phys. C - Particles and Fields  {\bf 49}, 485-490 (1991)
    \bibitem{ZFITTER}
            Computer code by D. Bardin et al., Z. Phys. C - Particles and
            Fields  {\bf 44}, 493 (1989), Nucl. Phys.
            {\bf B351} 1 (1991); Phys. Lett. {\bf B255} (1991) and CERN-TH
            6443/92 (May 1992).
    \bibitem{lynn}
            B.W.Lynn, High-precision tests of electroweak physics on the
            $Z^0$ resonance, Proceedings of the Workshop
            on Polarization at LEP, CERN 88-06, Sept. 1988, ed.
            G.Alexander at al., Vol. 1, p.24. at LEP1,
\end{thebibliography}
\end{document}